\documentclass[aps,prb,superscriptaddress,twocolumn]{revtex4}

\usepackage{graphicx}
\usepackage{times}
\usepackage{natbib}
\usepackage{color}
\usepackage{amsfonts}
\usepackage{amsmath}

\usepackage[colorlinks,bookmarks=false,citecolor=blue,linkcolor=blue,urlcolor=blue]{hyperref}

\begin{document}


\title{Non-equilibrium hysteresis and spin relaxation in the mixed-anisotropy dipolar coupled spin-glass LiHo$_{0.5}$Er$_{0.5}$F$_{4}$}


\author{J. O. Piatek}
\email[]{julian.piatek@epfl.ch}

\author{I. Kovacevic}

\author{P. Babkevich}

\author{B. Dalla Piazza}

\author{S. Neithardt}
 
\affiliation{LQM, ICMP, Ecole Polytechnique F\'{e}d\'{e}rale de Lausanne, CH-1015 Lausanne, Switzerland}

\author{J. Gavilano}
\affiliation{Laboratory for Neutron Scattering, Paul Scherrer Institut, CH-5232 Villigen, Switzerland}

\author{K. W. Kr\"{a}mer}
\affiliation{Department of Chemistry and Biochemistry, University of Bern, CH-3012 Bern, Switzerland}

\author{H. M. R\o{}nnow}

\affiliation{LQM, ICMP, Ecole Polytechnique F\'{e}d\'{e}rale de Lausanne, CH-1015 Lausanne, Switzerland}
\affiliation{RIKEN Centre for Emergent Matter Science (CEMS), Wako 351-0198, Japan}

\date{\today}
\begin{abstract}
We present a study of the model spin-glass LiHo$_{0.5}$Er$_{0.5}$F$_4$ using simultaneous AC susceptibility, magnetization and magnetocaloric effect measurements along with small angle neutron scattering (SANS) at sub-Kelvin temperatures. All measured bulk quantities reveal hysteretic behavior when the field is applied along the crystallographic c axis. Furthermore avalanche-like relaxation is observed in a static field after ramping from the zero-field-cooled state up to $200 - 300$~Oe. SANS measurements are employed to track the microscopic spin reconfiguration throughout both the hysteresis loop and the related relaxation. Comparing the SANS data to inhomogeneous mean-field calculations performed on a box of one million unit cells provides a real-space picture of the spin configuration. We discover that the avalanche is being driven by released Zeeman energy, which heats the sample and creates positive feedback, continuing the avalanche. The combination of SANS and mean-field simulations reveal that the conventional distribution of cluster sizes is replaced by one with a depletion of intermediate cluster sizes for much of the hysteresis loop.
\end{abstract}


\maketitle

Since the discovery of spin-glasses (SGs), considerable research has been dedicated to understanding their peculiar dynamical properties, where spins freeze out and respond to external stimuli with a characteristic time which can range from picoseconds to hours \cite{mydosh1993}. While the majority of zero-field equilibrium behavior is well established~\cite{Ogielski1985,levy_critical_1988, Fischer1991}, hysteresis and non-equilibrium properties remain active areas of research~\cite{young_spin_1997,vincent_slow_1997}. 

Perhaps the most intriguing zero-field non-equilibrium effect studied in detail is that of aging, rejuvenation and memory~\cite{jonason_memory_1998,dupuis_aging_2001}, which manifest themselves depending on the thermal history of the SG as it is cooled. Typical SGs also show non-equilibrium relaxation in both the thermoremanent magnetization, the magnetization acquired when field-cooled (FC), and the isothermal remanent magnetization, the instantaneous magnetization obtained by applying a field following zero-field-cooling (ZFC)~\cite{bert_spin_2004,zotev_role_2003}. Examples of more exotic behavior include the frequency dependent effects observed in LiHo$_{0.045}$Y$_{0.955}$F$_4$~\cite{Ghosh2002}, which are only seen when the sample is weakly coupled to the thermal bath~\cite{Quilliam2008,schmidt_using_2014}.

LiHo$_{0.5}$Er$_{0.5}$F$_4$ exhibits a low-temperature SG state below $T_{\text{g}}\sim0.4~\text{K}$,  which can be interpreted as co-existence of both Ising and $XY$ SGs~\cite{piatek_phase_2013}. The bulk of the magnetic properties appear to come from the Ising Ho$^{3+}$ spins which, according to neutron scattering, form small clusters highly elongated along the Ising axis and show no sign of long-range order (LRO). The combination of a  well characterized Hamiltonian, high quality samples and the ability to control the level of frustration make this an ideal system for studying SG behavior.

Here we present simultaneously measured AC susceptibility $\chi_{AC}$, magnetization $\mathbf{M}$ and magnetocaloric effect (MCE) measurements and complementary small angle neutron scattering (SANS) in the presence of a small field applied along the c axis. Hysteretic behavior is seen in all measured quantities and an avalanche-like relaxation is observed in the field range of 200-300 Oe. Inhomogeneous Mean-Field (iMF) calculations carried out on a box of one million unit cells qualitatively reproduce magnetization and SANS data, giving insight into real-space spin configurations.

\begin{figure}[htb]
\includegraphics[width=1\columnwidth]{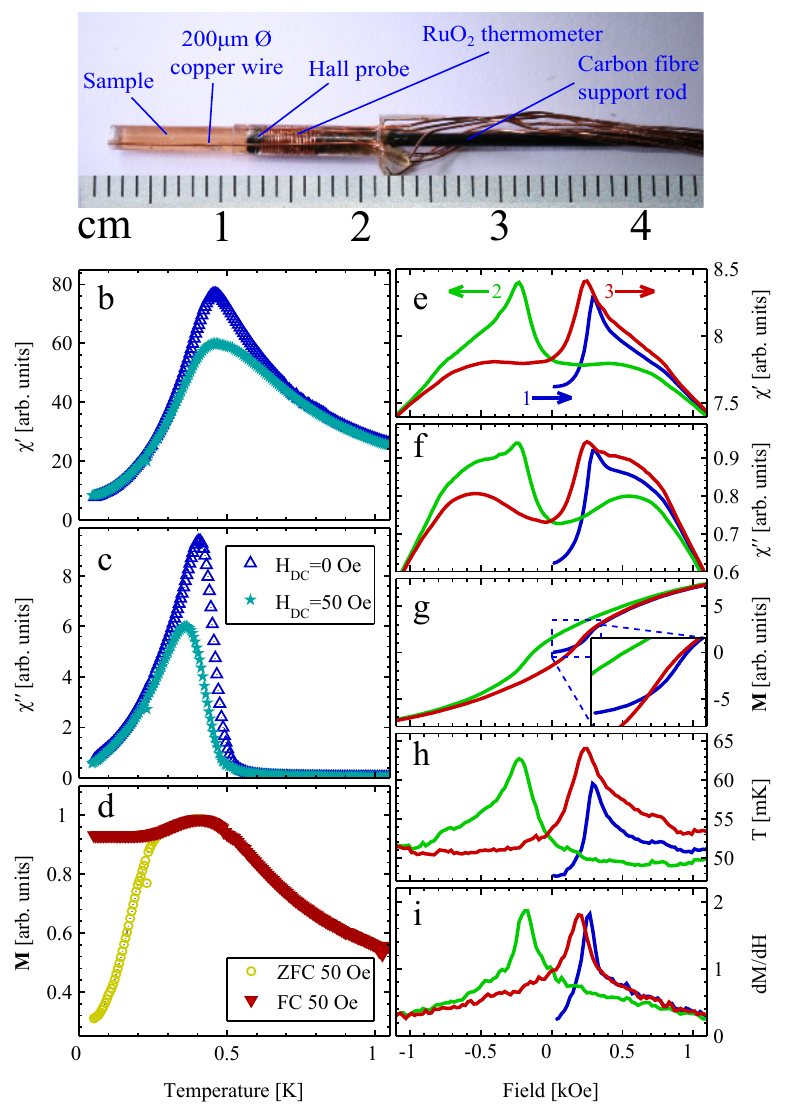}
\caption{(Color online). (a) Picture of sample. (b,c) $\chi^{\prime}\left(T\right)$ and $\chi^{\prime\prime}\left(T\right)$ in 0 \& 50 Oe. (d) ZFC \& FC $\mathbf M\left(T\right)$ in 50 Oe. (e-i) Hysteresis loops of $\chi^\prime$, $\chi^{\prime\prime}$, $\mathbf M$, $T_{sample}
$ and $\mathrm{d} M/\mathrm{d} H$.\label{bulk}}
\end{figure}

Bulk properties, $\chi_{AC}$, $\mathbf{M}$, $T$, were measured simultaneously on a $1.2 \times 1.2 \times 10$ mm single crystal with the length along the crystallographic c axis. As mentioned, spin-glass (SG) properties depend on the thermal and field path, so measuring all quantities simultaneously offers clear advantages.  A small ChenYang CYSJ166A GaAs hall effect sensor to measure $\mathbf M$ and a bare Vishay 4.7~k$\Omega$ RuO$_2$ resistor to measure $T$ have been glued on to the top of the crystal. The resulting package has been encapsulated in Stycast W19 and thermally anchored with four 200$~\mu \text{m}$ copper wires embedded in the Stycast as shown in Fig.~\ref{bulk} (a). The wires are attached to the mixing chamber of an Oxford Instruments Kelvinox dilution fridge which is placed inside a 9 T superconducting magnet. The AC susceptibility has been measured in a mutual inductance susceptometer with an AC field of 42 mOe oscillating at 545 Hz. The magnetization has been measured with a lock-in amplifier using a 1$~\mu$A Hall current oscillating at 77 Hz.

The temperature dependent complex susceptibility in zero field and 50 Oe is shown in Fig.~\ref{bulk} (b,c). ZFC and FC temperature dependent magnetization in 50 Oe is shown in panel (d). The ZFC-FC splitting in $\mathbf M \left( T \right)$ and the peak in $\chi\left( T\right)$, which is suppressed rapidly in a field, are consistent with the reported mixed Ising-$XY$ SG phase~\cite{piatek_phase_2013}. Panels (e-i) show the central region of a hysteresis-loop measuring $\chi_{AC}$ (e,f), $\mathbf M$ (g), $T_{\text{sample}}$ (h) and $\mathrm{d} M/\mathrm{d} H$ (i). Starting in the ZFC state the field is swept at 10 Oe/min from 0 to 2.5 kOe ($P_0$), from 2.5 to -2.5 kOe ($P_\downarrow$) and from -2.5 to 2.5 kOe ($P_\uparrow$) while maintaining the mixing chamber at $50\pm0.1$ mK.

Focusing first on $\chi_{AC}\left(\mathbf H\right)$, a hysteretic peak is observed in both $\chi^{\prime}$ and $\chi^{\prime\prime}$ along with a qualitative difference between $P_0$ and $P_\uparrow$. The sample temperature shows similar behavior to the AC susceptibility. The DC susceptibility ($\mathrm{d} M/\mathrm{d} H$) is much sharper than that of the AC susceptibility, which has shoulders just above the peak field of 290 Oe. The magnetization shows a hysteresis which is typical of a SG state, a very narrow hysteresis loop which has a distinctive `S' shape.

\begin{figure}[tb]
\includegraphics[width=1\columnwidth]{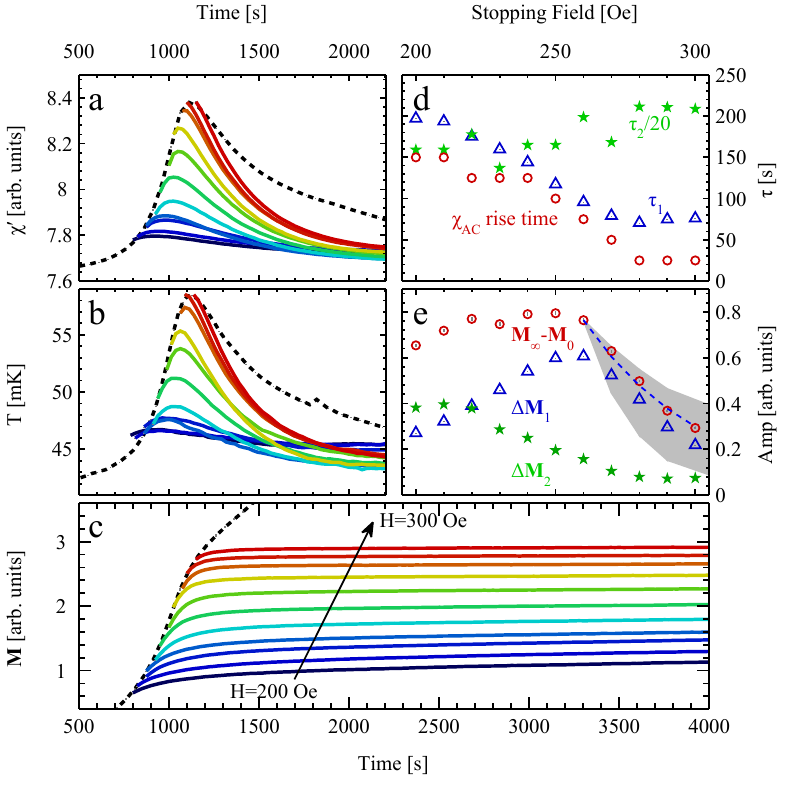}
\caption{(Color online). (a-c) Relaxation of $\chi^\prime$, $T_{sample}$ and $\mathbf M$ after constant ramps up to 200-300 Oe. Black dashed line shows a continuous field ramp. (d-e) Extracted time constants and amplitudes of fits of $\mathbf{M}$ to Eq.~\ref{double_exp}. In (e) the shaded region and dashed line shows an expected total relaxation assuming it began at $H = 260$ Oe.}
\label{avalanche}
\end{figure}

The SG dynamics were studied following field ramps at 16 Oe/min when starting from a ZFC state  using the following protocol. The sample is warmed to 1 K, thermalized for 5 minutes, cooled to base temperature at a ramp rate of 100 mK/min and subsequently thermalized for one hour. The field is then ramped at a rate of 16 Oe/min and when it reaches the desired value the relaxation is measured for two hours. All bulk properties are measured and checked for consistency throughout each step of the process to ensure there no major deviations from the desired $\left(T,\mathbf{H} \right)$ path. The relaxation of $\chi^\prime$, $T_{sample}$ and $\mathbf M$ as a function of time for stopping fields spaced every 10 Oe between 200 and 300 Oe is shown in Fig.~\ref{avalanche} (a-c). The black dashed line shows a continuous field ramp. 

Relaxation is observed in all quantities measured and is most dramatic in the temperature and susceptibility, which continue to increase for tens of seconds before relaxing for several thousand seconds. The magnetization relaxes upwards over the course of several hours.

Normally the relaxation of magnetization of a spin glass after application of a magnetic field change at a temperature deep in the spin glass phase occurs on a logarithmically extensive time scale \cite{nordblad_time_1986,sandlund_experimental_1988}, however the data recorded here are well described by two exponential decays of the form:

\begin{equation}
\mathbf M=\mathbf M_\infty - \Delta \mathbf M_1\exp^{-\frac{t}{\tau_1}}-\Delta \mathbf M_2\exp^{-\frac{t}{\tau_2}},
\label{double_exp}
\end{equation}

where $\mathbf M_\infty$ is the final value of the magnetization after relaxation, $\Delta \mathbf  M_1$ and $\Delta \mathbf  M_2$ are the amplitudes of the fast and slow exponential relaxations, and $\tau_1$ and $\tau_2$ their time constants. Thus while the magnetization relaxation does not reach equilibrium within the experimental time scale, extrapolating the exponential fits allows to estimate the equilibrium magnetization $\mathbf M_\infty$.

The dependence of the time constants and amplitudes extracted from the fits are shown in Fig.~\ref{avalanche} (d,e). $\tau_2$ remains relatively constant while $\tau_1$ gradually decreases as field increases. The time for which the susceptibility continues to rise after the field is stopped ($\chi_{AC}$ rise time) is found to reflect the behavior of $\tau_1$ (panel d). The difference between the final and starting magnetizations, $\mathbf M_\infty$-$\mathbf M_0$ remains constant up to 260 Oe where it begins to drop off (panel e). It can be modeled by postulating that the rapid relaxation begins at 260 Oe, with a $\tau_1$ in the range extracted from fits to Eq.~\ref{double_exp} (gray region) and the best fit $\tau_1=160$~s (dashed line).

Aiming to obtain insight on the path-dependent spin configurations behind the hysteresis behavior, small angle neutron scattering (SANS) was performed on SANS-I at the Paul Scherrer Institute (PSI) in Villigen, Switzerland. The sample used was a 1~mm thick, 10~mm diameter circular disk, whose $ac$ plane coincides with the plane of the disk and is perpendicular to the neutron beam. The instrument has been used with the detector 3 m from the sample with a collimation of 4.5 m before the sample and $\lambda = 0.53$ nm. Each SANS image was measured for one minute as a trade off between statistics and rapid measurements required to study the non-equilibrium properties. The configuration was used in order to have the largest possible $\mathbf{Q}$ range, which was constrained by the small magnet window. This technique allows us to probe the distinctive butterfly-shaped scattering which is observed around ferromagnetic Bragg peak (BP) positions~\cite{piatek_phase_2013} due to short range correlations in this dipolar coupled system. 

\begin{figure}[tbp]
\includegraphics[width=1\columnwidth]{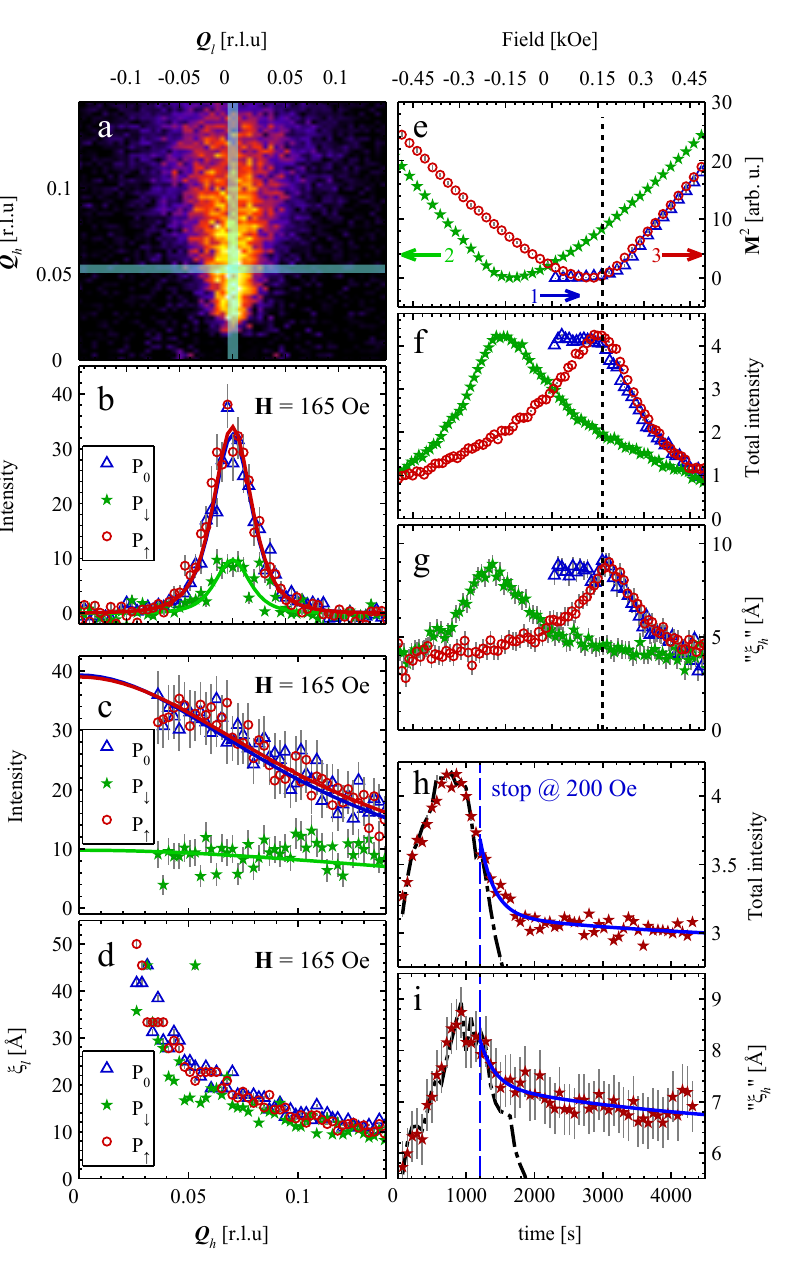}
\caption{(Color online). (a) Symmetrized SANS image, at T=50 mK, H=0 Oe, indicating cuts used in (b,c). (d) $\xi_l$ as a function of $\mathbf{Q}_h$. Data is taken from an image at $\mathbf H=165$~Oe with blue triangles indicating the $P_0$, green stars the $P_\downarrow$ and red circles the $P_\uparrow$ legs of a hysteresis loop. (e-f) Hysteresis scan of  $\mathbf M^2$, total SANS intensity and $\xi_h$ of scattering. (h-i) Relaxation and fit equivalent to Eq.~\ref{double_exp} of total SANS intensity and $\xi_h$ after ramping to 200 Oe (black dashed line shows continuous field-ramp).}
\label{sans}
\end{figure}

A typical SANS image, symmetrized around $\mathbf{Q}_h=0$, is shown Fig.~\ref{sans} (a).  The green shaded regions indicate those used for $\mathbf{Q}_l$ and $\mathbf{Q}_h$ cuts (b,c) which have been fit with a combination of a Lorentzian squared and Lorentzian (LSL) line shape, appropriate for systems with random fields~\cite{birgeneau_random-field_1983,lovesey_diffraction_1984,richards_spin-glass_1984}. The line shape is found to be predominately Lorentzian along the $\mathbf{Q}_h$ direction and Lorentzian squared along $\mathbf{Q}_l$. Cuts along $\mathbf{Q}_l$ symmetric peaks are fit to extract the $\mathbf{Q}_h$ dependence of $\xi_l$ (d). Panels (e-g) show the hysteresis loop of $\mathbf M^2$ (from Fig.~\ref{bulk} (g)), SANS integrated intensity and $``\xi_h"$~\footnote{While $\mathbf{Q}_h$ can be fit when peaked, it becomes flat for a large range of fields, incompatible with the definition of $\xi$. Therefore $``\xi_h"$ in Fig.~\ref{sans} (g,i) only partially reflects a correlation length in the conventional sense.}.

Relaxation measurements analogous to those in Fig.~\ref{avalanche} have been performed at 200 Oe and are shown in Fig.~\ref{sans} (h,i). The continuous field ramp is indicated by the dashed black line. Once the field is stopped a very clear relaxation is observed which is well described by a similar double exponential to that in Eq.~\ref{double_exp}, using the values of $\tau_1$ and $\tau_2$ extracted from the 200 Oe $\mathbf M(t)$ relaxation.

To complement bulk and SANS data, we developed an extension of mean-field theory dubbed inhomogeneous Mean-Field (iMF) which includes disorder effects. We start with the full Hamiltonian

\begin{eqnarray} {\mathcal H}&=&\sum_i\Big[{\mathcal
	H}_{\text{CF}}^{}({\bf J}_i^{})+ A {\bf J}_i^{}\cdot{\bf
	I}_i^{}-g\mu_B^{}{\bf J}_i^{}\cdot{\bf H}\Big]\nonumber
\\&-&{\textstyle\frac{1}{2}}\sum_{ij}
\sum_{\alpha\beta}{\mathcal
	J}_D^{}D_{\alpha\beta}^{}(ij)\mathbf{J}_{i\alpha}^{}\mathbf{J}_{j\beta}^{}
-{\textstyle\frac{1}{2}}\sum_{ij}^{n.n.}{\mathcal J}_{12}^{}\,{\bf
	J}_i\cdot{\bf J}_j^{} \quad
\end{eqnarray}

where the terms are respectively the crystal field, hyperfine interaction, Zeeman term, Dipole coupling and nearest neighbor interaction. The crystal-field and nearest-neighbor parameters have been determined previously for both LiHoF$_4$~\cite{ronnow_magnetic_2007} and LiErF$_4$~\cite{kraemer_dipolar_2012}. The mean-field approximation consists of replacing $ \mathbf{J}_{i} \cdot \mathbf{J}_{j}$ with $\mathbf{J}_{i} \cdot \left\langle \mathbf{J}_{j} \right\rangle$ and then solving the Hamiltonian using self-consistency equations. In contrast with regular virtual-crystal MF theory for mixed systems, the iMF approach starts out by generating a random realization of the Ho-Er disordered mixture on a large finite-size lattice. The mean-field self-consistency equations are then iteratively solved for each site in parallel. iMF calculations have been carried out using NVIDIA GPUs on a box containing a million unit cells. A GPU is used as the highly parallelizable dipole sum is the most time consuming segment of the code. Using a NVIDIA GTX 760 results in a reduction of calculation time by a factor of $\sim 20$ when compared to an Intel core i5 3350p CPU.
  
The result of each iteration of the iMF self-consistency is similar to how the physical system freezes. By performing the iMF calculations in series, where for each $\left(T, \mathbf{H} \right)$ point the mean-field state obtained in the previous point is used as an initial state, path-dependent configurations can be created. This allows the simulation of meta-stable states, which can then be inspected for distribution of clusters and correlations.

\begin{figure}[tbp]
\includegraphics[width=1\columnwidth]{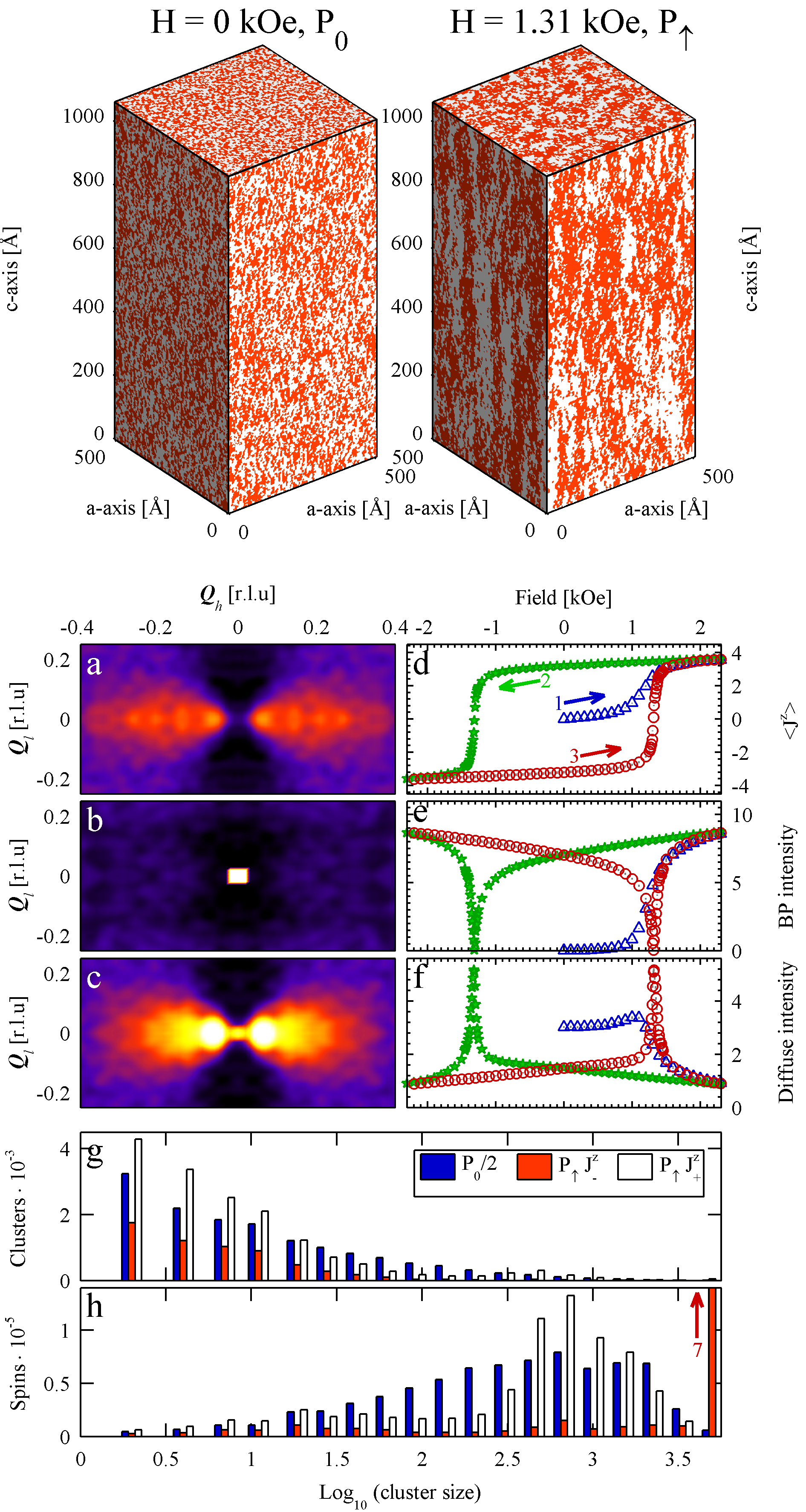}
\caption{(Color online). iMF simulations of LiHo$_{0.5}$Er$_{0.5}$F$_4$. (Top) real-space configuration of Ho$^{3+}$ moments at the ZFC state and 1.31 kOe $P_\uparrow$. SANS intensity of Ho$^{3+}$ moments in (a) ZFC state, (b) 1.31 kOe on $P_\downarrow$ and (c) $P_\uparrow$. (d-f) Hysteresis scan showing $\left\langle J^{z} \right\rangle$, intensity of BP and diffuse scattering. (g) Histogram of distribution of the cluster sizes and (h) number of spins in those clusters.}
\label{iMF}
\end{figure}

The top of Fig.~\ref{iMF} illustrates how two very different zero magnetization states can be reached using respectively the ZFC state (left) or the $\mathbf H=1.31$~kOe $P_\uparrow$ state (right). The direction of the Ho$^{3+}$ moments is represented with white pixels for $\left\langle J^z \right\rangle =5.5$ and red pixels for $\left\langle J^z\right\rangle =-5.5$ revealing a much more uniform distribution in the ZFC state. 

To compare these real-space configurations with SANS data, we calculate $\mathcal{S}\left(\mathbf{Q}\right)$ by Fourier transforming the real space configuration, applying a 5 pixel 3D smoothing using an 1.5 pixel FWHM Gaussian filter and then taking the square. The Bragg peak (BP) intensity is calculated by summing the central 5x5 pixels to remove possible finite-size effects~\footnote{If a finite system with periodic boundary conditions forms long range order with \textit{e.g.} two domains, the calculated BP will split into two separated by one pixel}. The diffuse intensity is calculated by summing over all remaining pixels in the image with $\left| \mathbf Q \right|<0.5$~[r.l.u].
 
Images of $\mathcal{S}\left(\mathbf{Q}\right)$ of the ZFC state, the $P_\downarrow$ and $P_\uparrow$ states at $H=1.31$~kOe are shown in Fig.~\ref{iMF} (a-c), which qualitatively reproduce the measured SANS image (\textit{c.f.} Fig.~\ref{sans} (a)). The full hysteresis loop of $\left\langle J^z \right\rangle$, BP intensity and diffuse scattering is shown in Fig.~\ref{iMF} (d-f), which correspond to $\mathbf{M}$ in Fig.~\ref{bulk} (g), and $\mathbf{M}^2$ and total intensity in Fig.~\ref{sans} (e,f) respectively. The qualitative agreement between $\mathcal{S}\left(\mathbf{Q}\right)$ in theory and experiment and their behavior as a function of field leads us to believe that the real-space spin configurations calculated are also representative of those in the experimental system and merit further attention. 

To better quantify the difference in the  $\left\langle J^z\right\rangle=0 $ configurations, the 2D sizes of clusters present in the $ab$ plane are calculated. The analysis can be restricted to this 2D plane as $\xi_l$ as a function of $\left(\mathbf{Q}_h\right)$ is independent of path (Fig.~\ref{sans} (d)). This means that regardless of other changes in the clusters, their aspect ratio remains constant. Cluster sizes are calculated by picking a starting pixel and recursively counting its adjacent pixels with the same polarization. The number and size of clusters has been counted for 40 horizontal cuts and has been used to generate the histograms of cluster size shown in Fig.~\ref{iMF} (g,h). The top histogram shows the number of clusters as a function of $\log_{10}\left(\mathrm{cluster\ size}\right)$ and the bottom one shows the number of spins contained within these clusters.

Having presented bulk, SANS and simulation results, we now interpret them. There are several peculiarities in $\mathbf M \left(\mathbf H\right)$. First, $\mathbf M\left(P_0\right)$ does not lie within the global hysteresis loop as one would expect~\cite{katzgraber_finite_2007}. Second, the system saturates at a low field of 2.5 kOe, in contrast to other SG systems which require very large fields to saturate~\cite{mydosh1993}. Finally, the sudden jump in magnetization around 250 Oe is not typically seen in a SG, but has been predicted~\cite{yoshino_step-wise_2007} for a mesoscopic SG. A similar jump has been observed in the magnetization of Fe$_x$Mn$_{1-x}$TiO$_3$~\cite{ito_magnetization_1997}, albeit without the anomalously flat ZFC magnetization before the jump which is observed here. Magnetization reversal jumps have also been observed in many canonical spin-glasses in FC hysteresis loops~\cite{monod_magnetic_1979,bert_spin_2004,retat_susceptibility_1983,uehara_staircase_1986}, although jumps are not seen in $P_0$ of a ZFC hysteresis loop.

The close resemblance of Fig.~\ref{bulk} (h,i) explains the MCE as due to Zeeman energy being released as spins are flipped. The power released by the Zeeman term can be expressed as $P\propto \mathrm{d} M/\mathrm{d} H \cdot \mathrm{d}H/\mathrm{dt}$ and therefore $P\propto \mathrm{d}M/\mathrm{d}H$ since the field ramp rate is constant. Assuming specific heat and thermal conductivity are independent of the $\sim10$ mK temperature variations, the released Zeeman power translates into a temperature change $\Delta T(H) \propto \mathrm{d}M/\mathrm{d}H$. This temperature change is what drives most of the response in $\chi_{AC}\left(H\right)$. Taking $\chi_{AC}\left(T\right)$ from Fig.~\ref{bulk}(b,c) we can conclude that the electronic moments are about 10 mK warmer than the thermometer while qualitatively behaving identically~\footnote{At 50mK $\chi^\prime(T)$ = 7.7 arb.u. with a slope of 0.03 arb.u./mK.}.

The combination of the MCE and the two-exponential decay of the magnetization gives a plausible model of the avalanche-like relaxation. As the field is stopped, the magnetization continues to increase rapidly, releasing Zeeman energy and heating the sample. As the sample warms, the SG becomes more dynamic, allowing for magnetization to grow even faster.  The self-driven heating continues for tens of seconds, before the sample is no longer releasing energy faster than it is being cooled, and is followed by more typical SG relaxation~\cite{smith_relaxation_1994}. In this scenario the former process is the origin of $\tau_1$ while the latter is that of $\tau_2$. An interesting question deserving theoretical scrutiny is whether this heating is uniform throughout the sample or if it acts locally to cause only near-neighbors to unfreeze.

A possible explanation as to why such dynamics occur here and have not been observed in other SGs is the combination of poor thermal conductivity, large specific heat and non-equilibrium thermally activated dynamics due to large anisotropies in LiHo$_{x}$M$_{1-x}$F$_4$ SGs~\cite{barbara_activated_2007,gingras_collective_2011}. The former two qualities make any temperature relaxation of the sample much more pronounced. The latter has been suggested as a possible explanation for the observed rounding of the linear and non-linear susceptibilities around $T_\text{f}$ in LiHo$_x$Y$_{1-x}$F$_4$~\cite{Jonsson2007}, although the effect deep within the frozen state remains unclear. 

The SANS measurement complemented by iMF calculations can be used to understand the rearrangement of the spins during a hysteresis loop. The ZFC state consists of a conventional distribution of all cluster sizes, resulting in a LSL line shape of $\mathcal{S}\left(\mathbf{Q}\right)$ and shows no LRO~\cite{piatek_phase_2013}. When a field is applied, first $\xi$ increases, implying larger clusters being created until above $H=165$~Oe both $\xi$ and diffuse intensity decrease, as intensity moves into the FM BP.  This is confirmed by comparing to $\mathbf M^2$, which for LRO is approximately proportional to the BP intensity~\footnote{Although LRO in a domain aligned against $\mathbf M$ will add to the BP but subtract from $\mathbf M^2$}. The hysteresis in intensity implies that the majority of the spins remain LRO until the field is reversed. This picture is qualitatively supported by the iMF calculations.

When reducing $\mathbf H$ from the high-field state, $\mathcal{S}\left(\mathbf{Q}_h\right)$ increases in intensity but remains flat (see $P_\downarrow$ in Fig.~\ref{sans} (c)). This is a clear indication that a conventional distribution of cluster sizes is no longer present. The dependence of $\xi_l\left(\mathbf Q_h\right)$ demonstrates that while the distribution of clusters is altered their aspect ratio remains the same. Using this information, it is then possible to restrict further analysis of the distribution of clusters to the $ab$ plane, as has been done to generate the histograms in Fig.~\ref{iMF} (g,h), revealing several features of the unconventional cluster distribution. First, most unflipped  moments remain percolated throughout the system in a single cluster. Second, the majority of flipped clusters are very small, containing less than tens of spins. Third, compared to the ZFC state, there is a depletion of clusters containing several tens to hundreds of spins.

To conclude we have observed two previously undocumented phenomena in SG systems. The first is the upward relaxation of magnetization immediately following a field ramp from the ZFC state. The relaxation is believed to be driven by flipping spins releasing their Zeeman energy, which causes an `avalanche' where the energy released allows for even more spins to flip. This process happens for several tens of seconds and can be seen in every measured quantity. Subsequently, by coupling SANS and iMF we have gained insight into the unconventional microscopic distribution of cluster sizes along the hysteretic paths.

We gratefully acknowledge help in the SANS experiment from K. Prsa which was performed at the Swiss spallation neutron source (SINQ) of the Paul Scherrer Institute (PSI) in Switzerland. This work is funded by the Swiss SNF and the European Research Council CONQUEST project.

\end{document}